\documentclass[aps,PRL,reprint,showpacs,superscriptaddress,
footinbib,citeautoscript]{revtex4-1}

\usepackage{graphicx}

\usepackage[utf8]{inputenc}
\usepackage{csquotes}
\usepackage[american]{babel}
\usepackage[T1]{fontenc}
\usepackage{enumerate}
\usepackage{mdwlist}
\usepackage[activate=normal]{pdfcprot}
\usepackage{bbding}
\usepackage{color}
\usepackage{amsmath}
\usepackage{eqnarray,amsmath}
\usepackage{amssymb}
\usepackage{amsmath}
\usepackage{amsfonts}
\usepackage{mathrsfs}
\usepackage[titletoc,title]{appendix}
\newcommand{\specialcell}[2][c]{\begin{tabular}[#1]{@{}l@{}}#2\end{tabular}}
\usepackage{bm}
\usepackage{dcolumn}
\usepackage{color}
\usepackage[normalem]{ulem}
\usepackage[colorlinks=true,citecolor=blue]{hyperref}
\hypersetup{colorlinks=true,citecolor=blue,linkcolor=red
,urlcolor=blue}

\newcommand{\beq}{\begin{equation}}
\newcommand{\eeq}{\end{equation}}
\newcommand{\beqa}{\begin{eqnarray}}
\newcommand{\eeqa}{\end{eqnarray}}

\begin{document}

\title{Anomalous plasmon mode in strained Weyl semimetals}
\author{Shiva Heidari}
\affiliation{School of Physics, Institute for Research in Fundamental Sciences, IPM, Tehran, 19395-5531, Iran}
\author{Dimitrie Culcer}
\affiliation{School  of  Physics,  University  of  New  South  Wales,  Kensington,  NSW  2052,  Australia}
\affiliation{ARC Centre of Excellence in Future Low-Energy Electronics Technologies,  UNSW Node,  Sydney 2052,  Australia}
\author{ Reza Asgari}
\email{asgari@ipm.ir}
\affiliation{School of Physics, Institute for Research in Fundamental Sciences, IPM, Tehran, 19395-5531, Iran}
\affiliation{ARC Centre of Excellence in Future Low-Energy Electronics Technologies,  UNSW Node,  Sydney 2052,  Australia}
\begin{abstract}
An exotic anomalous plasmon mode is found in strained Weyl semimetals utilizing the topological Landau Fermi liquid and chiral kinetic theories, in which quasiparticle interactions are modeled by long-range Coulomb and residual short-range interactions. The gapped collective mode is derived from the dynamical charge pumping between the bulk and the surface and behaves like $k_{\rm F}^{-1}$. The charge oscillations are accurately determined by the coupling between the induced electric field and the background pseudofields. This  novel mode unidirectionally disperses along the pseudomagnetic field and manifests itself in an unusual thermal conductivity in apparent violation of the Wiedemann-Franz law. The excitation can be achieved experimentally by mechanical vibrations of the crystal lattice in the THz regime.
\end{abstract}
\maketitle

\section{Introduction}
 Collective excitations in systems with long-range Coulomb interactions are referred to as longitudinal bulk plasmons \citep{PhysRev.92.609}. They are consistent with the classical plasma picture and can be controlled by tailoring the spatial region filled by a charged plasma. Plasmonics is based on interaction processes between electromagnetic radiation and itinerant charges. It seamlessly combines fundamental research and applications across areas ranging from condensed matter physics~\citep{Tame2013} to compact stars~\citep{Kouveliotou_1999} and plasma in the early universe~\citep{Durrer2013}, to color engineering, chemistry, biology and medicine~\citep{shahbazyan2013plasmonics}.  

Dirac and Weyl materials mimic the properties of high-energy relativistic matter and provide an excellent opportunity to explore novel quantum effects~\citep{RevModPhys.90.015001}. Their topological band structure and electron correlations are accurately described by topological Fermi liquid theory \citep{Son_2012, PhysRevB.95.205113, PhysRevLett.93.206602, PhysRevB.95.205113}. The non-zero Berry phase of quasiparticles in Weyl semimetals together with the novel axionic term in the electromagnetic response \citep{PhysRevLett.111.027201, PhysRevB.86.115133, Landsteiner_2016} make the dynamics of excitations completely different from collective modes in ordinary metals. A considerable effort is devoted to identifying novel excitations in interacting Weyl fermions in a three-dimensional (3D) relativistic-like plasma, which may originate from anomaly-induced intra- or inter-chiral particle number fluctuations.

One example is the violation of axial current conservation, termed the chiral~\citep{Note1} anomaly, i.e. $\partial_\mu J_5^\mu=\frac{e^2}{3\pi^2} \bm{E}\cdot \bm{B}$ \citep{Nielsen_1983,PhysRevB.86.115133}, stemming from the topological modification of the electromagnetic response \citep{RevModPhys.90.015001,PhysRevB.86.115133}. It leads to a non-dissipative current along a magnetic field through the chiral magnetic effect in the presence of an axial chemical potential. The collective dynamics of Weyl fermions in the presence of quantum anomalies undergoes a qualitatively change in the dispersion of conventional collective modes \citep{PhysRevB.92.201407,PhysRevB.99.075104,PhysRevB.91.035114,Gorsky_2013,PhysRevB.98.165122,Sukhachov_2018,Ahn_2016} and even gives rise to novel and unprecedented types of excitations \citep{Song_2019,Gorbar_2017,PhysRevD.91.125014,PhysRevB.99.075104,PhysRevB.91.035114,Sukhachov_2018, Landsteiner_2016}. 
Moreover, the electron-phonon coupling in strained Weyl semimetals in the form of elastic gauge fields, $\bm{{\cal A}}^{el}$, \citep{PhysRevLett.115.177202,PhysRevX.6.041021,Ilan_2019} leads to new collective dynamics \citep{PhysRevResearch.1.032040,PhysRevResearch.1.033070,Gorbar_2017}. Notably, the phonon collective excitations receive considerable modifications in both the longitudinal \citep{PhysRevResearch.1.032040} and the optical \citep{PhysRevB.100.165427} branches due to electron-phonon interactions. The presence of both ordinary and strain-induced pseudofields, i.e. $\bm{{\cal E}}^{el}=\partial_t \bm{{\cal A}}^{el}$ and $\bm{{\cal B}}^{el}=\nabla \times \bm{{\cal A}}^{el}$, not only modifies the chiral anomaly equation, but also results in the non-conservation of local charges \citep{Grushin_2016,PhysRevX.6.041021,Ilan_2019,PhysRevB.87.235306}.

In this paper we identify a new anomalous plasmon (AP) mode in interacting type-I Weyl semimetals in the presence of a pseudomagnetic field induced by strain. 
We assume the conventional model of Weyl semimetals with the minimum number of two opposite-chirality nodes when time-reversal symmetry is broken \citep{Nielsen_1981}. We demonstrate that bulk charge oscillations induce an electric field that couples to the background pseudofields. This coupling leads to dynamical charge pumping between the bulk and the surface and vice versa through the apparent non-conservation of local charge in the bulk, i.e. $\partial_\mu J^\mu= \frac{e^2}{2\pi^2} \bm{{\tilde E}}{({\bm r},t)} \cdot \bm{{\cal B}}^{el}$. Hence, the AP mode is to be distinguished from chiral plasmons and magneto-plasmons in that charge fluctuations do not occur between the nodes, but between the bulk and the boundaries and are mediated by the Fermi arcs. Adopting the framework of topological Landau Fermi liquid theory including strain-induced pseudo-electromagnetic fields, we derive the $\bm{q}-$dependent plasmon dispersion stemming from anomalous electronic transport phenomena. Most importantly, this AP mode only carries a charge current and therefore, it is no longer a chiral mode. Note that Weyl fermions in tilted Weyl semimetals that emerge at the boundary between the electron and hole pockets (due to the Lorentz symmetry breaking) are completely different from standard type-I Weyl semimetals. Accordingly, the tilt effect on AP mode needs further discussion which is beyond the scope of this paper. 

In addition, the AP mode as a bosonic quasiparticle only disperses along the pseudomagnetic field and may manifest itself in an unusual thermal conductivity through the violation of the Wiedemann-Franz law. The AP can lead to remarkable thermodynamic phenomena, such as quantum oscillations in the thermal conductivity due to the pseudomagnetic field which can be considered as a smoking gun.  

This paper is organized as follows. We commence with a description of the topological Fermi liquid theory in Sec. \ref{sec:Theory}, followed by the details of the Berry curvature, the residual short-range and the long-range Coulomb interactions. The comprehensive discussions on collective dynamics and thermal properties of the system are reported. In conclusion, we summarize our main findings in Sec. \ref{sec:conclusion}.

\section{Topological Fermi liquid theory}\label{sec:Theory}
We consider a system in which the low-energy effective Hamiltonian in the continuum limit in the vicinity of the nodal points is given by 
$
{\cal H}=v_{\rm F}({\bm q}+\chi \bm{{\cal A}}^{el})\cdot {\bm \sigma}
$
where $\chi=\pm$ labels the chirality of the nodal points. Such nodes are connected by Fermi arc surface states and topologically stabilized against any slight perturbations regardless of symmetry \citep{Herring_1937}. Since the pseudomagnetic field plays the same role as a real magnetic field, it can modify the low-energy energy of Weyl fermions by a term owing to the orbital magnetic moment $\bm{m}_k^\chi$, i.e. $\epsilon^\chi(\bm{k})=v_{\rm F} k-\bm{{\cal B}}^{\chi}\cdot \bm{m}^\chi_{(k)}$ where $\bm{m}^\chi_{(k)}=-v_{\rm F} k \bm{\Omega}^\chi_{(k)}$ \citep{RevModPhys.82.1959}. Here, $v_{\rm F}$ is the Fermi velocity, $\bm{\Omega}^\chi=\chi \bm{\Omega}=\chi \hat{k}/2|k|^2$ is the isotropic Berry curvature and $\bm{B}^\chi=\chi \bm{{\cal B}}^{el}$ is the elastic-in-origin pseudomagnetic field that couples to the Weyl fermions oppositely in the two nodes at $\bm{b}$ and $-\bm{b}$. This orbital magnetic moment stems from the self rotation of the Bloch wave packet around its center \citep{RevModPhys.82.1959}.
 
The interaction-induced renormalized local quasiparticle energy is given by
$
\tilde{\epsilon}^\chi (\bm{k},\bm{r},t)=\epsilon^\chi (\bm{k}) + \delta \epsilon^\chi (\bm{k},\bm{r},t)$ where
$\epsilon^\chi(\bm{k})=v_{\rm F} k (1+ \bm{B}^\chi \cdot \bm{\Omega}^\chi)$ is the fermionic energy dispersion.  It is worth mentioning that $\epsilon^\chi(\bm{k})$ is modified by the contributions due to all filled electronic states through the Berry curvature. This is somewhat distinct from the Landau theory, which merely involves quasiparticles within a small range of $k_B T$ \citep{PhysRevB.95.205113,PhysRevLett.93.206602,PhysRevLett.109.162001}. Remarkably, $\epsilon^\chi (\bm{k})$ is independent of the specific nature of interactions and carries information on the topological characteristics of the band structure. 
The inhomogeneous part of energy due to the presence of the collective
mode and the intrinsic interactions is given by
\begin{equation} \label{eq4}
\delta \epsilon^\chi=  \sum_{\chi^\prime} \int \dfrac{d\bm{k}^\prime}{(2\pi)^3} {\cal D}_{(\hat{k}^\prime)} \{{\cal F}_{\chi \chi^\prime} (\bm{k},\bm{k}^\prime)+v_{\bm{q}}\}\delta f_{\chi^\prime}.
\end{equation}
It takes account of both ${\cal F}_{\chi \chi^\prime} (\bm{k},\bm{k}^\prime)$ and $v_{\bm{q}}=e^2/ \epsilon_0 {\bm{q}}^2$ as a residual short-range interaction between two fermions of type $\chi$, $\chi^\prime$ and the long-range Coulomb interaction, respectively. The electronic fluctuation of the distribution function in the vicinity of chiral Fermi surfaces is given by
$
\delta f_\chi (\bm{k},\bm{r},t)=f_\chi (\bm{k},\bm{r},t)-f^{(eq)}_\chi (\bm{k})
$. We suppose quantum oscillations are sought in the form of plane waves with frequency $\omega$ and wave-vector $\bm{q}$, $\delta f_\chi (\bm{k},\bm{r},t)=\delta f_\chi (\bm{k}) e^{i(\bm{q} \cdot \bm{r} - \omega t)}$. The equilibrium distribution function $f^{(eq)}_\chi (\bm{k})=[e^{\beta (\epsilon_k^\chi-\mu_\chi)}+1]^{-1}$, where $\beta=(k_B T)^{-1}$ and $\mu_\chi=\mu^{(eq)}+\chi \mu_5$ is the effective chemical potential for the right- and left- handed fermions. For $\bm{k}$ and $\bm{k}^\prime$ near the Fermi surface where $\epsilon_{\bm{k}}=\epsilon_{\bm{k}^\prime}=\epsilon_F$, the interaction term ${\cal F}_{\chi \chi^\prime} (\bm{k},\bm{k}^\prime)$ depends only on the angle between the direction of $\bm{k}$, $\bm{k}^\prime$ and on the chiralities $\chi$ and $\chi^\prime$.
The factor ${\cal D}_{(\hat{k})}=1-\bm{\Omega} \cdot \bm{{\cal B}}^{el}$ ensures the phase space modification satisfies Liouville's theorem \citep{PhysRevLett.95.137204}.

\subsection{Collective dynamics}
The topologically modified semiclassical Boltzmann formalism can be embedded in the framework of chiral kinetic theory. The collective dynamics of a pair of chiral Fermi surfaces are described by the time evolution of quasiparticle distribution function, which satisfies
\begin{eqnarray} \label{eq}
\partial_t f_{\chi} (\bm{k},\bm{r},t) & + (\dot{\bm{r}}_\chi \cdot \bm{\nabla}_{\bm{r}}+\dot{\bm{k}}_\chi \cdot \bm{\nabla}_{\bm{k}}) f_\chi (\bm{k},\bm{r},t) \\ & =\bm{{\cal I}} (\delta f^\chi (\bm{k},\bm{r},t)).\nonumber
\end{eqnarray}
Scattering processes are accounted for by the collision integral on the RHS. The semiclassical equations of motion in topological Fermi liquid theory in the absence of time reversal symmetry read
\begin{eqnarray}
& {\cal D}_{(\hat{k})} \dot{\bm{k}}_\chi= \bm{E}^{\chi}-\bm{\nabla}_{\bm{r}} \tilde{\epsilon}(\bm{k},\bm{r},t)+ \bm{\nabla}_{\bm{k}} \tilde{\epsilon}(\bm{k},\bm{r},t) \times \bm{B}^\chi \nonumber\\ & - [\bm{\nabla}_r \tilde{\epsilon}(\bm{k},\bm{r},t) \cdot \bm{B}^\chi] \bm{\Omega}^\chi (\bm{k})-(\bm{E}^{\chi} \cdot \bm{B}^{\chi}) \bm{\Omega}^\chi (k), \nonumber\\ \\
&{\cal D}_{(\hat{k})} \dot{\bm{r}}_\chi= \bm{\nabla}_{\bm{k}} \tilde{\epsilon}(\bm{k},\bm{r},t) - \bm{\nabla}_{\bm{r}} \tilde{\epsilon}(\bm{k},\bm{r},t) \times \bm{\Omega}^\chi (\bm{k}) \nonumber\\ & + [\bm{\nabla}_{\bm{k}} \tilde{\epsilon}(\bm{k},\bm{r},t) \cdot \bm{\Omega}^\chi (\bm{k})] \bm{B}^\chi-\bm{E}^{\chi} \times \bm{\Omega}^\chi(k),\nonumber
\end{eqnarray}
where $\bm{E}^\chi=\chi \bm{{\cal E}}^{el}$. The topological Landau Fermi liquid theory is valid in the semiclassical regime, where $v_{\rm F}\sqrt{{\cal B}^{el}} \ll \tau^{-1} \ll \mu$. Here, $\tau$ is the quasiparticle lifetime. The semiclassical limit ensures $|\mu| \gg E_{n=1}(k=0)$, where $E_{n=1}(k=0)$ is the dispersion of first level generated by strain-induced pseudomagnetic field \citep{PhysRevX.6.041021,PhysRevLett.119.075301}. As a result, many Landau-levels have been occupied and then we can ignore pseudo-Landau-level quantization. Worth mentioning that the semiclassical regime dictates that only intraband transitions have dominated the process, and any transitions with frequencies lower than $2 \mu$ are prohibited due to the Pauli blocking.

The dynamical equation describing the quantum oscillation of excited quasiparticle with $(\bm{k},\chi)$ interacting with fermions of type $(\bm{k}^\prime,\chi^\prime)$ is given by
\begin{eqnarray} \label{eq11}
& - i \omega {\cal D}_{(\hat{k})} \delta f_{\chi}(\bm{k})+i \bm{q} \cdot [\bm{v}_k+ (\bm{v}_k \cdot \bm{\Omega}^\chi (\bm{k})) \bm{B}^\chi] \delta f_\chi (\bm{k})+\nonumber\\ & (- \partial f^{(eq)}/\partial \epsilon_k )\bm{v}_k \cdot \sum_{\chi^\prime , k^\prime} \bm{\Pi}^{\chi,\chi^\prime}_{(\bm{q},\bm{k},\bm{k}^\prime)} +(\bm{v}_k \times \bm{B}^\chi ) \cdot \bm{\nabla}_k \delta f_\chi (\bm{k}) \nonumber\\ &- \bm{v}_k \cdot [\bm{E}^{\chi}+(\bm{E}^{\chi} \cdot \bm{B}^{\chi}) \bm{\Omega}^\chi (k)] ={\cal D}_{(\hat{k})} \bm{{\cal I}} (\delta f_\chi (\bm{k})),
\end{eqnarray}
The modified interaction-induced drag force is
\begin{equation}
\bm{\Pi}^{\chi,\chi^\prime}_{(\bm{q},\bm{k},\bm{k}^\prime)}= i {\cal D}_{(\hat{k}^\prime)} [{\cal F}_{\chi \chi^\prime}(\xi)+v_{\bf q}] [\bm{q} +( \bm{q} \cdot \bm{B}^\chi) \bm{\Omega}^\chi (\bm{k})] \delta f_{\chi^\prime} (\bm{k}^\prime).\nonumber
\end{equation}
We assume that steady state processes are accurately described by the relaxation time approximation, $ d f^\chi(\bm{k},\bm{r},t)/d t=-(f_\chi (\bm{k},\bm{r},t)-f_\chi ^{(eq)}(\bm{k}))/ \tau(\bm{k}) $ valid for elastic impurity-scattering process  when the scattering centers are homogeneously distributed and the linear response regime is assumed \citep{PhysRevB.89.195137,PhysRevB.90.165115}. We also consider, for simplicity, a $\bm{k}$-independent relaxation time, i.e. $\tau(\bm{k})\rightarrow\tau $, since all processes occur close to the Fermi momentum and thus $\bm{k}$-dependent relaxation time does not affect on results. Collision-induced quantum oscillation of particles could be decomposed into a thermal relaxation time $\tau_{th}$, an inter-node relaxation time $\tau_c$, and the relaxation time-scale of the charge-density imbalance between the bulk and the boundaries denoted by $\tau_a$ and provided by the process in which right-moving modes in the bulk scatter back to the left-moving modes near the boundaries. If the electrons have to travel the length $L$ to reach the surface, $N_a=(L/l_a)^2$ scattering events should be ocurred to take the route \citep{PhysRevX.6.041021}. Here, $l_a$ is the magnetic length almost equals to $l_a\simeq \sqrt{\hbar c/ e {\cal B}^{el}}\simeq 0.8 \ \mu$m for ${\cal B}^{el}\simeq 1$ mT. In samples with $L\gg l_a$ it turns out that $\tau_a / \tau_{th}\simeq (L/l_a)^2 \gg 1$, so we assume $\tau_{th} \ll \tau_{a}$ \citep{PhysRevX.6.041021}. Accordingly, we restrict our analysis to time-scales $\tau_{th} \ll \tau_a $ and $\tau_{th} \ll \tau_c $ where the latter denotes the well-known chiral limit. Such a chiral limit has been considered in other works \citep{Song_2019,Son_2012}. In this limit, the scattering rate $\tau_{th}^{-1}$ is fast enough to relax any deviation of the Fermi surface and then establish thermal equilibrium in each chiral Fermi surface.

Having carried out the integration over $\bm{k}$, Eq. (\ref{eq11}) can be correctly interpreted as the non-conservation of chiral charges in the presence of both ordinary and pseudofields $\partial_t \delta n_\chi+\nabla \cdot \bm{{\cal J}}_\chi=(\chi \bm{{\cal E}}^{el} \cdot \bm{{\cal B}}^{el}+\bm{{\tilde E}}{(\bm{r},t)} \cdot \bm{{\cal B}}^{el})/4 \pi^2$  (More details are presented in Appendix \ref{appa}).  Here, the intrinsic electric field $\bm{{\tilde E}}{(\bm{r},t)}$ induced by charge density oscillation, $\delta n(\bm{r},t)=\sum_{k,\chi} {\cal D}_{(\hat{k})} \delta f_\chi (\bm{k},\bm{r},t)$, has been extracted from $\bm{{\tilde E}}{(\bm{r},t)}=-e \nabla \varphi(\bm{r},t)$ where $\varphi(\bm{r},t)$ represent the dynamical scalar potential satisfying the Poisson equation $q^2 \varphi(\bm{r},t)=\frac{e}{\epsilon_0}\delta n(\bm{r},t)$. As we prove in Appendix \ref{appa}, subtracting and adding the charge and current associated with each node leads to the following covariant form of the \textit{novel chiral anomaly} and the \textit{non-conservation of local charge}, respectively;
\begin{subequations}
\begin{eqnarray}
&& \partial_\mu {\cal J}^\mu_5=\dfrac{e^2}{2 \pi^2} \bm{{\cal E}}^{el} \cdot \bm{{\cal B}}^{el}, \label{chiral-eq}\\
&& \partial_\mu {\cal J}^\mu_{(\bm{r},t)}=\dfrac{e^2}{2 \pi^2} \bm{{\tilde E}}{(\bm{r},t)} \cdot \bm{{\cal B}}^{el}, \label{total-eq}
\end{eqnarray}
\end{subequations}
where we define ${\cal J}^\mu_5=({\cal J}^\mu_+ -{\cal J}^\mu_-)$ and ${\cal J}^\mu=({\cal J}^\mu_+ +{\cal J}^\mu_-)$. The coupling between ordinary intrinsic  electric field to the pseudomagnetic field leads to a local charge non-conservation (Eq.~\ref{total-eq}), while the elastic pseudofields lead to a chiral anomaly (Eq.~\ref{chiral-eq}). The chiral anomaly in Eq.~(\ref{chiral-eq}) describes the strain-induced charge pumping between the nodes with opposite chirality, $\delta n_\chi^{(1)}= \chi\frac{e^2\tau_c}{2 \pi^2} (\bm{{\cal E}^{el}}\cdot \bm{{\cal B}^{el}})$, leads to a slight shift in chemical potential as $\delta \mu_\chi^{(1)}= \chi (\bm{{\cal E}^{el}}\cdot \bm{{\cal B}^{el}})\tau_c /2 $. It is noted that this strain-induced chirality imbalance is independent of the dynamics of collective excitations and it stems from the extrinsic elastic pseudofields. On the other hand, the coupling between pseudomagnetic field and the induced electric field $\bm{{\tilde E}}{(\bm{r},t)}$ owing to the charge dynamics leads to an unexpected local charge non-conservation, Eq.~(\ref{total-eq}), that can be interpreted as a charge pumping between the bulk and the boundaries of the system \citep{PhysRevX.6.041021,Landsteiner_2016}. Although the result of Eq.~(\ref{total-eq}) seems to be unphysical, considering both the bulk and the surface contribution can restore the charge conservation \citep{Landsteiner_2016,PhysRevB.89.075124}. The violation of local charge conservation in Weyl semimetals naturally arises from the fact that the current conservation equation only includes the bulk region, hence the excess charge is expected to come from the edge of the system \citep{PhysRevX.6.041021,Ilan_2019,Grushin_2016}. This non-conservation problem can be circumvented by adding the so-called Bardeen-Zumino polynomial, ${\cal J}^\mu \rightarrow {\cal J}^\mu+\frac{e^2}{2\pi^2} \bm{{\cal A}}^{el} \times \bm{\tilde{E}}$ to the electric current, which renders the consistent version of the anomaly equation, $\partial_\mu {\cal J}^\mu=0$~\citep{Bardeen_1986,Landsteiner_2016,Gorbar_2017}.
\begin{figure}[t]
\includegraphics[scale=0.5]{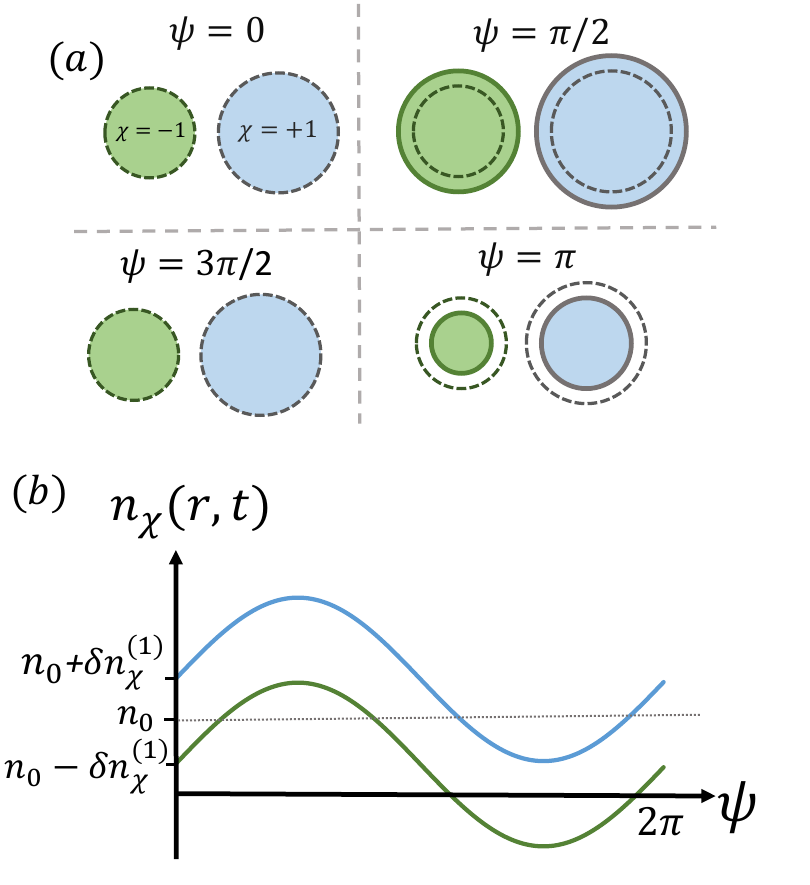}
\caption{\label{fig1} (Color online) (a): Colored disks represent the coherent breathing-like fluctuation of the bulk Fermi surfaces for a single pair of Weyl nodes owing to the anomaly-induced charge transfer between the bulk and the surface. The dashed circles denote the equilibrium position of the Fermi surfaces $\mu^{(eq)}_\chi=\mu+ \chi |\delta \mu_\chi^{(1)}|$. (b): The blue and green lines denote the charge density creation and annihilation when $\psi$ sweep along $[0,2\pi]$ and $\delta n^{(1)}_\chi$ represents the chirality imbalance between two nodes due to the term $\propto \bm{{\cal E}^{el}} \cdot \bm{{\cal B}^{el}}$ in the chiral anomaly equation. We define $\psi=\bm{q}\cdot \bm{r}-\omega t$ as a quantum phase of the collective motion of charges.} 
\end{figure}
The strain-induced local charge non-conservation in Weyl semimetals, Eq.~(\ref{total-eq}), plays a key role in driving a new collective mode. The charge density imbalance induced  between the bulk and the surface is 
\begin{equation} \label{density-anomaly}
\delta n_\chi^{(2)}=\dfrac{e^2 \tau_a}{2 \pi^2} \bm{{\tilde E}} \cdot \bm{{\cal B}}^{el} e^{i(\bm{q}\cdot \bm{r}-\omega t)}. 
\end{equation}
The induced charge in the bulk is distributed among all the empty states above $\epsilon_F$ and leads to a slight shift in the chemical potential
$\delta \mu_\chi^{(2)}(\bm{r},t) \approx \frac{e^2 \tau_a}{\mu^2} \bm{{\tilde E}} \cdot \bm{{\cal B}}^{el}  e^{i(\bm{q}\cdot \bm{r}-\omega t)},
$
extracted from the semiclassical formalism where pseudo-Landau level quantization is unimportant [Appendix~\ref{appb}]. The sign of the anomaly-induced charge density in the bulk depends on the phase of the charge fluctuation $\psi=\bm{q}\cdot \bm{r}-\omega t$, i.e. $\delta n_\chi^{(2)}>0$ for $\psi=\pi/2$, $\delta n_\chi^{(2)}<0$ for $\psi=3\pi/2$ and $\delta n_\chi^{(2)}=0$ for $\psi=0,\pi$ with respect to the $\mu=\mu_0+\chi |\delta \mu^{(1)}|$. Figure \ref{fig1} demonstrates contributions from both chiral anomaly, $\delta n_\chi^{(1)}$, and the dynamics of collective excitation, $\delta n_\chi^{(2)}(t)$. The total charge density of a single node with chirality $\chi$ is given by $n_\chi(t)=n_0+\delta n_\chi(t)$, where $n_0$ is the intrinsic charge density and  $\delta n_\chi(t)=\delta n_\chi^{(1)}+\delta n_\chi^{(2)}(t)$ is the anomaly-induced charge density above each Fermi surface. It means charges are pumped from the bulk to the surface and vice versa through $\delta n_\chi^{(2)}(t)$, then it gives rise to the coherent breathing-like fluctuation on the bulk Fermi surface.

By making use of the modified anomaly equation $\partial_t \delta n_\chi+\nabla \cdot \bm{{\cal J}}_\chi=(\chi \bm{{\cal E}}^{el} \cdot {\bm{{\cal B}}^{el}})/4 \pi^2$ incorporating the boundary contributions and the current expressions, we investigate the spectrum of AP excitations. For the sake of simplicity, we consider ${\cal F}_{\chi,\chi^\prime}(\hat{k},\hat{k}^\prime)={\cal F}_0$ as a constant and valid for small Fermi surfaces, and also $\epsilon_k \rightarrow \epsilon_k+\chi |\delta \mu_\chi^{(1)}|$. After straightforward calculations we obtain
\begin{equation}
\delta n_\chi=\dfrac{ ({\cal F}_0+\dfrac{e^2}{\epsilon_0 q^2})\sum_{\chi^\prime} \chi^\prime \delta n_{\chi^\prime}}{(\omega+i\tau^{-1})-\chi \bm{q} \cdot \bm{\alpha}/2\pi^2}~\frac{\bm{q} \cdot \bm{\alpha}}{ 2 \pi^2},
\end{equation}
where $\bm{\alpha}={\cal B}^{el}/2 k_F^2$. Assuming a finite frequency $\omega$, vanishingly small scattering rate, i.e. $\tau_{c,a}^{-1}\rightarrow 0$, and small $\bm{q} \cdot \bm{\alpha}$ we get the following spectral equation for the collective mode 
\begin{eqnarray}
&& 1=\sum_\chi  \dfrac{\chi k_F^2 \bm{q} \cdot \bm{\alpha}/2 \pi^2}{\omega-\chi \bm{q} \cdot \bm{\alpha}/2 \pi^2}({\cal F}_0+\dfrac{e^2}{\epsilon_0 q^2})= \nonumber\\ &&\sum_\chi \dfrac{\chi  k_F^2\bm{q} \cdot \bm{\alpha} }{2 \pi^2 \omega} (1+ \dfrac{\chi \bm{q} \cdot \bm{\alpha}}{\omega}+\cdots)({\cal F}_0+\dfrac{e^2}{\epsilon_0 q^2})\simeq\nonumber\\ && (a_1 + a_2 |q|^2) (\dfrac{\hat{q}\cdot \bm{\alpha}}{\omega})^2 + (a_1 + a_2 |q|^2) |q|^2 (\dfrac{\hat{q}\cdot \bm{\alpha}}{\omega})^4+\cdots ,
\end{eqnarray}
where $a_1=e^2 k_F^2/ 2\epsilon_0\pi^2$  and $a_2={\cal F}_0 k_F^2/2 \pi^2$. The corresponding AP frequency in the long-wavelength limit would be
\begin{equation}
\omega^{AP}(q)= |\hat{q} \cdot \bm{\alpha}| \sqrt{a_1+(1+a_2) |q|^2}.
\end{equation}
It is worth noting that this mode only disperses along the pseudomagnetic field and is only tied to the local charge oscillations, but the inter-node chiral fluctuations, on the other hand, would be absent. 
The plasmon gap is obtained by keeping terms up to the zeroth order of $\bm{q}$: 
\begin{equation}
\omega_{\bm{q} \rightarrow 0}^{AP}=\sqrt{\dfrac{e^2}{2 \epsilon_0 \pi^2}} \dfrac{|\hat{q} \cdot \bm{{\cal B}}^{el} |}{2 k_F}.
\end{equation}
where the plasmon mode is proportional to $1/k_{\rm F}$. To reliably estimate the plasmon gap, we set $v_{\rm F}=2 \times 10^5$ m/s \citep{Lee_2015}, $\epsilon_F=100$ meV and $|{\cal B}^{el}|=1$ mT which gives $\omega_{\bm{q} \rightarrow 0}^{AP} \simeq 15$ THz. Such a low pseudomagnetic field can be generated by applying an infinitesimal in-plane lattice distortion with a twist angle $\Omega=1^\circ$. In this case, the deviation of lattice sites from their equilibrium positions is given by a vector $\bm{u}=\Omega \frac{z}{L} (\hat{z} \times \bm{r})$, where $\bm{r}$ is the position vector of each site and $L$ is the crystal length. The corresponding induced pseudomagnetic field associated to elastic gauge field of the form $A^{el}\simeq \Phi_0 \beta u_{ij} b^j$ would be ${\cal B}^{el}=\beta \Phi_0 b_3 \frac{\Omega}{4 L} \hat{z}$, where the strain tensor is defined as $u_{ij}=\frac{1}{2}(\partial_i u_j+\partial_j u_i)$. Recalling that $\beta \approx 2$ (Gr\"{u}neisen parameter), $\Phi_0=\hbar c/e \approx 6.5 \times 10^4 \ T$\AA$^2$ and $b_3=0.015$ \AA$^{-1}$ (half of the node separation in TaAs with lattice constant $a=3.45$ \AA  \citep{Xu_2015,Lv_2015}) and crystal length $L=1 \ \mu$m, it turns out that ${\cal B}^{el} \sim 1$ mT. Having mentioned before, such a small pseudomagnetic field can induce plasmon mode providing the building blocks for terahertz optical devices.
 
In the chiral limit, i.e. $\tau_c \gg \tau_{th}$, any deviation of the Fermi surface will be immediately washed out by a strong intra-node scattering process, therefore the only remaining collective mode is the gapped AP mode, which propagates along the pseudomagnetic field. The plasmon mode of 3D pristine Weyl semimetals in the absence of real electromagnetic fields is given by $\omega_p(1-\frac{1}{8\mu^2}q^2(1+F(2\mu,\omega_p))$ where $\omega_p=\sqrt{8 e^2 \mu^2/(3\pi\epsilon_0)}$, $F(x,y)=(x^4y^2-3x^6/5)/(y^2(x^2-y^2)^2)$ and $\mu=\hbar v_{\rm F} k_{\rm F}$ is the chemical potential~\cite{PhysRevB.91.035114}. Consequently, the strain-induced pseudo-magnetic field drives plasmon collective dynamics by generating charge fluctuations between the bulk and the boundaries without a background real magnetic field. 

\subsection{Thermal properties}
\begin{figure}[t]
\includegraphics[scale=1.1]{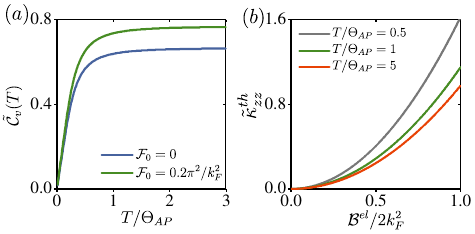}
\caption{\label{fig2}(Color online) (a): The renormalized specific heat $\tilde{{\cal C}}_v=(4\pi^2 (1+a_2)^{3/2}/k_B \Lambda^3){\cal C}_v$ as a function of temperature. The temperature is plotted in the units of Debye temperature for $\Theta_{AP}=\alpha \Lambda/ k_B$, where $\alpha={\cal B}^{el}/2 k_F^2$. The active presence of repulsive short-range interaction ${\cal F}_0$, generates the specific heat to increase especially in the higher temperature regions. We set here $a_1=\Lambda^2$ and $a_2=0,0.1$ for the blue and the green lines, respectively. (b): The renormalized thermal conductivity $\tilde{\kappa}_{zz}^{th}=(4 \pi^2/k_B \tau_p \Lambda^3 (1+a_2)^{1/2}) \kappa_{zz}^{th}$ as a function of pseudomagnetic field ${\cal B}^{el}$ for temperatures $T/\Theta_{AP}=0.5,1,5$ and $a_1=\Lambda^2$ and $a_2=1$.}
\end{figure}

The existence of an independent AP mode can be regarded as a bosonic quasiparticle in the Weyl Fermi liquid system. Such a strain-induced collective excitation may make a contribution to the thermal properties such as the specific heat and thermal conductivity. The total energy carried by the collective mode is defined as
$
{\cal U}=\sum_{\bm{q}} \omega_{q} \ {\cal G}^{(0)}(\bm{q},\bm{r},t)
$
, where ${\cal G}^{(0)}(\bm{q},\bm{r},t)=(e^{\beta \omega_{q}}-1)^{-1}$ is the equilibrium Bose-Einstein distribution function and $\omega_q$ is the dispersion of the collective mode. The specific heat, i.e. ${\cal C}_v=\partial {\cal U}/\partial t$, can be obtained as
\begin{equation}
{\cal C}_v(T)=k_B \sum_{|\bm{q}| < \Lambda} \dfrac{(\beta \omega_{q})^2}{4 \sinh^2 (\beta \omega_{q}/2)}
\end{equation}
where we consider $\Lambda$ as an ultraviolet cutoff for the wave vector integrals. We estimate $\Lambda\sim 1/a$ where $a$ is the lattice parameter. We consider a pseudomagnetic field parallel to the $z$-axis, and the dependence of the specific heat with respect to temperature is presented in Fig.~\ref{fig2}(a).

The specific heat behavior in the two limits of sufficiently low $T \ll \Theta_{AP}$ and high temperatures $T \gg \Theta_{AP}$ is given by
\begin{equation}
{\cal C}_v= \begin{cases}
\frac{k_B \Lambda^3}{6 (a_2+1)^{3/2} } B(a_1,a_2)\frac{T}{\Theta_{AP}}  \ \ \ &T \ll \Theta_{AP}\\\\
\dfrac{k_B \Lambda^3}{6 \pi^2}  &T \gg \Theta_{AP}
\end{cases}\;,\nonumber
\end{equation}
where $\Theta_{AP}=\alpha \Lambda / k_B$ represents the corresponding Debye temperature for APs and $B(a_1,a_2)=\sqrt{(a_2+1)(a_1+a_2+1)}/2-\frac{a_1}{2}\ln|(\sqrt{a_2+1}+\sqrt{a_1+a_2+1})/\sqrt{a_1}|$. For more details see Appendix \ref{appc}.

The thermal conductivity, $\kappa^{th}$, on the other hand, is defined via the heat current $\bm{j}^{th}=\kappa^{th} (- \nabla T)$. The thermal current associated with the unidirectional AP mode in terms of its spectrum is
\begin{equation} \label{th-curr}
\bm{j}^{th}=\sum_{|\bm{q}| < \Lambda} \omega_q (\nabla_q \omega_q) \ \delta {\cal G}(\bm{q},\bm{r},t).
\end{equation}
Here, $\delta {\cal G}(\bm{q},\bm{r},t)$ represents the stationary solution of Boltzmann equation, $\partial_{t}{\cal G}(\bm{q},\bm{r},t)=0$, and is given by 
\begin{eqnarray} \label{th-dist}
\delta {\cal G}(\bm{q},\bm{r},t)= &&{\cal G}(\bm{q},\bm{r},t)-{\cal G}^{(0)}(\bm{q},\bm{r},t)=-\tau_p(T) \bm{\dot{r}} \cdot \nabla_r {\cal G}^{(0)} \nonumber\\
=&& \tau_p(T) \dfrac{\alpha^2 (1+a_2)k_B \beta^2}{4 \sinh^2(\beta \omega_q/2)} |q| (-\nabla_z T),
\end{eqnarray}
where $\tau_p(T)$ is the relaxation time of the APs. Using Eqs. (\ref{th-curr}) and (\ref{th-dist}), the thermal conductivity can be expressed as follows;
\begin{equation}
\kappa_{zz}^{th}=\tau_p(T)k_B (1+a_2)^2\alpha^4 \beta^2 \sum_{|\bm{q}| < \Lambda} \dfrac{q^2}{ 4\sinh^{2}(\beta \omega_q/2)}.
\end{equation}
Figure \ref{fig2}(b) represents the thermal conductivity along the direction of AP propagation. The thermal conductivity for low-temperatures $T \ll \Theta_{AP}$ is proportional to $\kappa_{zz}^{th} \propto ({\cal B}^{el})^2 (\Theta_{AP}/T)$ and is temperature-independent at $T \gg \Theta_{AP}$, i.e. $\kappa_{zz}^{th} \propto ({\cal B}^{el})^2$ [Appendix.\ref{appc}]. The electronic conductivity, on the other hand, displays a Drude-like response in the dc regime~\cite{PhysRevLett.108.046602}, therefore, the presence of the unidirectional AP mode leads to an unusual and anisotropic thermal conductivity which violates the Wiedemann-Franz law. Such a violation of the Wiedemann-Franz law by the anomaly-induced chiral zero sound (CZS) mode in Weyl semimetals has been also theoretically proposed \citep{Song_2019} and experimentally confirmed \citep{PhysRevX.9.031036}, recently.

Ultimately, it should be mentioned that, in principle, the band bending effect may become important when the Fermi level lies well above the Van Hove (VH) energy, i.e. $\mu \gg \epsilon_{VH}$ \citep{PhysRevB.94.241105}. In this regime, two Weyl cones merge into an associated dispersion, and the notion of a chiral fermion is lost. For a minimal Hamiltonian including band bending, the Van Hove singularity happens around $\sim 0.5$ eV for a typical Weyl semimetal \citep{PhysRevB.101.085202}. For chemical potentials much smaller than $\epsilon_{VH}$, which is compatible with our Weyl system, the band bending effect is unimportant. Accordingly, the general conclusions do not change qualitatively under the assumption that we are below the corresponding Van Hove singularity.

\subsection{Anomalous plasmon mode in comparison with other collective modes in Weyl semimetals}

Since several collective modes of Weyl semimetals have been reported in various situations, a proper comparison with their results seems to be in order.
To commence with, we describe the collective modes of 3D normal Fermi liquid systems. In the presence of the long-range Coulomb interaction, charge collective mode occurs in the system in the absence of any external fields. The gapped mode behaves like $\omega=\omega^{(3D)}_p[1+\frac{9q^2}{10\kappa^2_3}]$ where $\omega^{(3D)}_p=\sqrt{4\pi e^2/m}$ and $\kappa_3=3\omega^{(3D)}_p/v^2_{\rm F}$. This mode originates from the intra-node fluctuations of the Fermi surface and finite net charge density is propagated by the plasmons. In Fermi liquid systems, a zero sound mode \citep{pines2018theory}, on the other hand, emerges in the presence of the residual short-range interaction. The zero sound disperses like $\omega=c_s q$ where the velocity is $c_s \propto \sqrt{F_0}$.

In the context of noninteracting Weyl semimetals, Gorsky and Zayakin \citep{Gorsky_2013} showed that the anomalous term in the current modifies the structure of the zero sound mode in the presence of a magnetic field. Jeong and Kim \citep{PhysRevB.98.165122} found the zero sound mode of Fermi surface fluctuations in a residual interacting Weyl metal phase in the presence of external electromagnetic fields. Gorbar et al. \citep{Gorbar_2017} proposed the chiral plasma mode in Weyl materials in constant magnetic and pseudomagnetic fields, taking into account the effects of dynamical electromagnetism. Stephanov et al. \citep{PhysRevD.91.125014} showed that the chiral magnetic wave emerges in the hydrodynamic regime; at frequencies smaller than the collision relaxation rate. Moreover, the chiral magnetic wave velocity is only determined by thermodynamic properties.
Chernodub and Vozmediano \citep{PhysRevResearch.1.032040} proposed the chiral sound wave in a strained wire of a Weyl semimetal which is a longitudinal charge density wave analog to the chiral magnetic wave driven by an elastic axial pseudomagnetic field.

In the context of interacting Weyl semimetals, on the other hand, the plasmon dispersion is distinct from that of conventional 3D metals \citep{lv2013dielectric} in the absence of external electromagnetic fields. Zhou et al. \citep{PhysRevB.91.035114} investigated the chiral anomaly effect on the charged plasmon mode within the random phase approximation. The long-range Coulomb interaction between electrons was considered.
Song and Dai \citep{Song_2019} proposed a chiral zero sound in Weyl semimetals under the magnetic field and in the presence of a residual short-range interaction. The sound velocity of chiral zero sound is proportional to the field strength in the weak field limit, whereas it oscillates dramatically in the strong field limit. The comprehensive results of the collective mode in Weyl semimetals are summarized in Table I in Appendix \ref{appd}.

The conclusion of these detailed comparisons is that we found an exotic anomalous plasmon mode in strained and interacting Weyl semimetals where quasiparticle interactions are modeled by the long-range Coulomb interaction and the residual short-range interaction. The new collective mode is derived from the dynamical charge pumping between the bulk and the surface and behaves like $k_{\rm F}^{-1}$. This novel mode unidirectionally disperses along the pseudomagnetic field.

\section{Conclusion}\label{sec:conclusion}
We have identified an anomalous plasmon mode as a novel type of cooperative motion of Weyl fermions in a distorted lattice as a unique signature of novel manifestation of the anomaly equations. Topological Fermi liquid theory with pseudofields is utilized to determine its exotic gapped dispersion relation. The AP mode only propagates along the pseudomagnetic field with a frequency of a few THz and vanishes in the absence of lattice distortion. This unidirectional mode is characterized by an oscillation of the charge density between the bulk and the boundaries triggered by the strain-induced anomalous non-conservation of local charge. The anomalous plasmon mode is completely different from other collective modes proposed for Weyl semimetals.

We have also shown the AP mode can lead to an unprecedented thermal conductivity along the pseudomagnetic field which does not satisfy the Wiedemann-Franz law. Such exotic thermal transport may be considered as strong evidence in experiments to confirm the existence of the AP mode.


\section{Acknowledgment}
DC and RA were supported by the Australian Research Council Centre of Excellence in Future Low-Energy Electronics Technologies (project number CE170100039).

\begin{widetext}
\appendix
\section{Topological Fermi Liquid theory and anomalous plasmon mode} \label{appa}
The time evolution of quasiparticle distribution function in the semiclassical limit is given by
\begin{equation} \label{eq}
\partial_t f_{\chi} (\bm{k},\bm{r},t) + (\dot{\bm{k}}_\chi \cdot \bm{\nabla}_{\bm{k}} + \dot{\bm{r}}_\chi \cdot \bm{\nabla}_{\bm{r}}) f_\chi (\bm{k},\bm{r},t) =\bm{{\cal I}} (\delta f^\chi (\bm{k},\bm{r},t)).
\end{equation}
The equations of motion typically receive necessary modifications due to the topological band structure, i.e. non-zero Berry curvature and the presence of pseudofields
\begin{equation} \label{eqs}
\begin{split}
& {\cal D}_{(\hat{k})} \dot{\bm{k}}_\chi= \bm{E}^{\chi}-\bm{\nabla}_{\bm{r}} \tilde{\epsilon}(\bm{k},\bm{r},t)+ \bm{\nabla}_{\bm{k}} \tilde{\epsilon}(\bm{k},\bm{r},t) \times \bm{B}^\chi - [\bm{\nabla}_r \tilde{\epsilon}(\bm{k},\bm{r},t) \cdot \bm{B}^\chi] \bm{\Omega}^\chi (\bm{k})-(\bm{E}^{\chi} \cdot \bm{B}^{\chi}) \bm{\Omega}^\chi (k), \\ \\
&{\cal D}_{(\hat{k})} \dot{\bm{r}}_\chi= \bm{\nabla}_{\bm{k}} \tilde{\epsilon}(\bm{k},\bm{r},t) - \bm{\nabla}_{\bm{r}} \tilde{\epsilon}(\bm{k},\bm{r},t) \times \bm{\Omega}^\chi (\bm{k}) + [\bm{\nabla}_{\bm{k}} \tilde{\epsilon}(\bm{k},\bm{r},t) \cdot \bm{\Omega}^\chi (\bm{k})] \bm{B}^\chi-\bm{E}^{\chi} \times \bm{\Omega}^\chi(k).
\end{split}
\end{equation}
Having replaced Eq. (\ref{eqs}) into Eq. (\ref{eq}), the dynamical equation of quasiparticles promptly becomes
\begin{equation}
\begin{split}
& - i \omega {\cal D}_{(\hat{k})} \delta f_{\chi}(\bm{k})+i \bm{q} \cdot \bm{v}_k \delta f_\chi (\bm{k})+(- \dfrac{\partial f^{(eq)}}{\partial \epsilon_k} ) i \bm{q} \cdot \bm{v}_k \sum_{\chi^\prime , k^\prime} {\cal D}_{(\hat{k}^\prime)} ({\cal F}_{\chi,\chi^\prime}(\xi)+\dfrac{e^2}{\epsilon_0 q^2})f_{\chi^\prime}(k^\prime)+(\bm{v}_k \times \bm{B}^\chi ) \cdot \bm{\nabla}_k \delta f_\chi (\bm{k})\\ & + i \bm{q} \cdot \bm{B}^\chi (- \dfrac{\partial f^{(eq)}}{\partial \epsilon_k}) \sum_{\chi^\prime , k^\prime} {\cal D}_{(\hat{k}^\prime)} ({\cal F}_{\chi,\chi^\prime}(\xi)+\dfrac{e^2}{\epsilon_0 q^2}) \bm{v}_k \cdot \bm{\Omega}^\chi (k) \delta f_{\chi^\prime}(k^\prime)+i \bm{q} \cdot \bm{B}^\chi (\bm{v}_k \cdot \bm{\Omega}^\chi (\bm{k})) \delta f_\chi (\bm{k})-\bm{v}_k \cdot \bm{E}^\chi \\ & - (\bm{E}^\chi \cdot \bm{B}^\chi)(\bm{v}_k\cdot \bm{\Omega}^\chi(k))={\cal D}_{(\hat{k})} {\cal I}(\delta f_\chi (k))
\end{split}
\end{equation}
It should be noted that terms of the form $E^\chi \delta f(k)$ are neglected due to the linear response consideration of $\delta f(k)$. Having carefully carried out the integration over $\bm{k}$ and using the fact that $\int d^3 k (\bm{v}_k \times \bm{B}^{\chi}) \cdot \nabla_k \delta f_\chi (k)=0$, $\int d^3 k (\bm{v}_k \cdot \bm{E}^\chi) =0 $, the above equation can be arranged and simplified in the following way;
\begin{equation} \label{eqdy}
\begin{split}
-i\omega \delta n_\chi+ i \bm{q} \cdot [\bm{{\cal J}}^{(0)}_\chi+\bm{{\cal J}}^{(1)}_\chi+\bm{{\cal J}}^{(2)}_\chi+\bm{{\cal J}}^{(3)}_\chi ]=& -i \bm{q} \cdot \bm{{\cal B}}^{el} \dfrac{e^2}{4 \pi^2 \epsilon_0 q^2} \sum_{\chi^\prime,k^\prime} {\cal D}_{(\hat{k}^\prime)} \delta f_{\chi^\prime}(\bm{k}^\prime,\bm{r},t)+\int \dfrac{d^3 k}{(2\pi)^3} (\bm{E}^\chi \cdot \bm{B}^\chi)(\bm{v}_k\cdot \bm{\Omega}^\chi(k)) \\ & +\int \dfrac{d^3 k}{(2\pi)^3} {\cal D}_{(\hat{k})} (-\dfrac{\delta f_\chi (k)}{\tau}),
\end{split}
\end{equation}
where we have reasonably assumed the relaxation time approximation for the collision process, i.e. ${\cal I}(\delta f_\chi (k))=(-\dfrac{\delta f_\chi (k)}{\tau})$, and the charge density and current contributions associated to chirality $\chi$ are defined as bellow (We set here $v_{\rm F}=\hbar=1$)
\begin{equation}
\delta n_\chi(r,t)=\int \dfrac{d^3 k}{(2\pi)^3} {\cal D}_{(\hat{k}^\prime)} \delta f_\chi (k,r,t)
\end{equation}
\begin{equation}
\bm{{\cal J}}^{(0)}_\chi(\bm{r},t)= \int \dfrac{d^3k}{(2\pi)^3} \bm{v}_k \delta f_\chi (\bm{k},\bm{r},t)
\end{equation}
\begin{equation}
\bm{{\cal J}}^{(1)}_\chi(\bm{r},t)=k_F^2 \int \dfrac{d \Gamma}{(2 \pi)^3} \hat{k} [\sum_{\chi^\prime,k^\prime} {\cal D}_{(\hat{k}^\prime)} {\cal F}_{\chi \chi^\prime}(\hat{k},\hat{k}^\prime) \delta f_{\chi^\prime}(\bm{k}^\prime,\bm{r},t)+\dfrac{e^2}{\epsilon_0 q^2} \delta n (\bm{r},t)]
\end{equation}
\begin{equation}
\bm{{\cal J}}^{(2)}_\chi(\bm{r},t)= \dfrac{\bm{{\cal B}^{el}}}{2} \int \dfrac{d \Gamma}{(2 \pi)^3} \sum_{\chi^\prime,k^\prime} {\cal D}_{(\hat{k}^\prime)} {\cal F}_{\chi \chi^\prime}(\hat{k},\hat{k}^\prime) \delta f_{\chi^\prime}(\bm{k}^\prime,\bm{r},t)
\end{equation}
\begin{equation}
\bm{{\cal J}}^{(3)}_\chi(\bm{r},t)=\int \dfrac{d^3k}{(2 \pi)^3} \dfrac{\bm{{\cal B}^{el}}}{2 |k|^2} \delta f_\chi (\bm{k},\bm{r},t)=\dfrac{\bm{{\cal B}}^{el}}{4\pi^2} \delta \mu_\chi(\bm{r},t)
\end{equation}
We assume that charge density oscillation $\delta n(r,t)$ satisfies the Poisson equation, $\nabla^2 \varphi(r,t)=-\frac{e}{\epsilon_0} \delta n(r,t)$, by defining $\varphi (r,t)$ as a dynamical scalar potential. The first term on the RHS of Eq. (\ref{eqdy}) can be expressed in terms of $\varphi (r,t)$ by using $\delta n(r,t)=\frac{\epsilon_0 q^2}{e} \varphi (r,t)$. By making use of  
$\varphi \nabla \cdot \bm{{\cal B}^{el}}=\nabla \cdot (\bm{{\cal B}^{el}} \varphi)- \bm{{\cal B}^{el}} \cdot \nabla \varphi$ we have
\begin{equation}
-i \bm{q} \cdot \bm{{\cal B}}^{el} \dfrac{e^2}{4 \pi^2 \epsilon_0 q^2} \delta n(r,t)=-i \dfrac{e \bm{q} \cdot \bm{{\cal B}}^{el}}{4 \pi^2}\varphi (r,t)=\dfrac{e^2 \bm{\tilde{E}}(r,t)\cdot \bm{{\cal B}}^{el}}{4 \pi^2}-\nabla \cdot [\bm{{\cal B}}^{el} \dfrac{e^2}{4 \pi^2\epsilon_0 q^2} \delta n(r,t)],
\end{equation}
where we have properly used $\bm{\tilde{E}}(r,t)=-e \nabla \varphi (r,t)$ or $\bm{\tilde{E}}(r,t)=-ie \bm{q} \varphi (r,t)$. Therefore, Eq. (\ref{eqdy}) can be simplified as
\begin{equation}
-i(\omega+i \tau^{-1}) \delta n_\chi+i \bm{q} \cdot \bm{{\cal J}}_\chi+i \bm{q} \cdot \bm{{\cal J}}^{(4)}_\chi=(\chi \bm{{\cal E}}^{el} \cdot \bm{{\cal B}}^{el}+\bm{\tilde{E}}{(\bm{r},t)} \cdot \bm{{\cal B}}^{el})/4 \pi^2,
\end{equation}
or in the form of the following continuity equation;
\begin{equation}
\partial_t \delta n_\chi+\nabla \cdot \bm{{\cal J}}_\chi+\nabla \cdot \bm{{\cal J}}^{(4)}_\chi=(\chi \bm{{\cal E}}^{el} \cdot \bm{{\cal B}}^{el}+\bm{\tilde{E}}{(\bm{r},t)} \cdot \bm{{\cal B}}^{el})/4 \pi^2.
\end{equation}
where
\begin{equation}
\bm{{\cal J}}^{(4)}_\chi(\bm{r},t)= \dfrac{\bm{{\cal B}^{el}}}{2} \int \dfrac{d \Gamma}{(2 \pi)^3} \sum_{\chi^\prime,k^\prime} {\cal D}_{(\hat{k}^\prime)} (\dfrac{e^2}{\epsilon_0 q^2}) \delta f_{\chi^\prime}(\bm{k}^\prime,\bm{r},t)=\bm{{\cal B}}^{el} \dfrac{e^2}{4\pi^2\epsilon_0 q^2} \delta n(r,t).
\end{equation}

Subtracting and adding the charge and current associated to each node leads to the following  \textit{covariant} form of the \textit{novel chiral anomaly} and the \textit{non-conservation of local charge}
\begin{subequations}
\begin{eqnarray}
&& \partial_\mu {\cal J}^\mu_5=\dfrac{e^2}{2 \pi^2} \bm{{\cal E}}^{el} \cdot \bm{{\cal B}}^{el}, \label{chiral-eq1}\\
&& \partial_\mu {\cal J}^\mu{(\bm{r},t)}=\dfrac{e^2}{2 \pi^2} \bm{\tilde{E}}{(\bm{r},t)} \cdot \bm{{\cal B}}^{el}, \label{total-eq1}
\end{eqnarray}
\end{subequations}
where $\bm{{\cal J}}{(\bm{r},t)}=\bm{{\cal J}}^{(0)}{(\bm{r},t)}+\bm{{\cal J}}^{(1)}{(\bm{r},t)}+\bm{{\cal J}}^{(2)}{(\bm{r},t)}+\bm{{\cal J}}^{(3)}{(\bm{r},t)}+\bm{{\cal J}}^{(4)}{(\bm{r},t)}$.

The chiral anomaly in Eq.~(\ref{chiral-eq1}) describes the strain-induced charge pumping between the nodes with opposite chirality leads to the chemical potential imbalance between two nodes.
Eq.~(\ref{total-eq1}), on the other hand, represents the violation of charge conservation. This non-conservation problem can be released by adding the so-called Bardeen-Zumino polynomial to the electric currant
\begin{equation}
j\rightarrow j+\delta j \ \, \ \ {\text where} \ \ \ \delta j=\dfrac{e^2}{2\pi^2} \bm{{\cal A}}^{el} \times \bm{\tilde{E}}{(\bm{r},t)}  \ \ \ \longrightarrow \partial_\mu (\delta j)=\dfrac{e^2}{2\pi^2} \bm{\tilde{E}}_{(\bm{r},t)}  \cdot \bm{{\cal B}}^{el}
\end{equation}
The modification of Eq.~(\ref{total-eq1}) renders the consistent version of the anomaly equation, $\partial_\mu {\cal J}^\mu=0$.

The Bardeen-Zumino polynomial which restores the non-conservation of local charge can be interpreted as the anomalous
current term which propagates from the bulk to the boundaries and vice versa through the Fermi arcs. The single node anomaly equation after applying the Bardeen correction would be 
\begin{equation} \label{app-ano}
\partial_t \delta n_\chi +\nabla \cdot \bm{{\cal J}}_\chi=(\chi \bm{{\cal E}}^{el}\cdot \bm{{\cal B}}^{el})/4\pi^2.
\end{equation}
Having utilized the above definition of current expression $\bm{{\cal J}}_\chi(r,t)$ and charge density fluctuation $\delta n_\chi (r,t)$, we can simply obtain Eq.~(\ref{app-ano}) in its Fourier form
\begin{equation}
(\omega+i\tau^{-1}) \delta n_\chi-\chi \dfrac{\bm{q}\cdot \bm{{\cal B}^{el}}}{4 \pi^2 k_F^2} \delta n_\chi=\dfrac{ \bm{q} \cdot \bm{{\cal B}^{el}}}{4 \pi^2} ({\cal F}_0+\dfrac{e^2}{\epsilon_0 q^2}) \sum_{\chi^\prime} \chi^\prime \delta n_{\chi^\prime},
\end{equation}
where the dynamics of charges are defined from the energy level $\chi |\delta \mu_\chi^{(1)}|$ in each node corresponds to a shift $\epsilon_k \rightarrow \epsilon_k+\chi |\delta \mu_\chi^{(1)}|$ in energy levels of charge fluctuations.
The above dynamical equation can be written in a more compact form
\begin{equation}
\delta n_\chi=\dfrac{ ({\cal F}_0+\dfrac{e^2}{\epsilon_0 q^2})\sum_{\chi^\prime} \chi^\prime \delta n_{\chi^\prime}}{(\omega+i\tau^{-1})-\chi \bm{q} \cdot \bm{\alpha}/2\pi^2}~\frac{\bm{q} \cdot \bm{\alpha}}{ 2 \pi^2},
\end{equation}
where $\bm{\alpha}={\cal B}^{el}/2 k_F^2$. Applying summation $\sum_{\chi} \chi$ to both sides and expand the denominator by assuming small $\bm{q} \cdot \alpha$ and finite frequency $\omega$  we reach the following polynomial equation
\begin{eqnarray}
 1=&& \sum_\chi  \dfrac{\chi k_F^2 \bm{q} \cdot \bm{\alpha}/2 \pi^2}{\omega-\chi \bm{q} \cdot \bm{\alpha}/2 \pi^2}({\cal F}_0+\dfrac{e^2}{\epsilon_0 q^2}) =\sum_\chi \dfrac{\chi  k_F^2\bm{q} \cdot \bm{\alpha} }{2 \pi^2 \omega} (1+ \dfrac{\chi \bm{q} \cdot \bm{\alpha}}{\omega}+\cdots)({\cal F}_0+\dfrac{e^2}{\epsilon_0 q^2})\nonumber\\ && \simeq (a_1 + a_2 |q|^2) (\dfrac{\hat{q}\cdot \bm{\alpha}}{\omega})^2 + (a_1 + a_2 |q|^2) |q|^2 (\dfrac{\hat{q}\cdot \bm{\alpha}}{\omega})^4+\cdots.
\end{eqnarray}
where $a_1=e^2 k_F^2/ 2\epsilon_0\pi^2$  and $a_2={\cal F}_0 k_F^2/2 \pi^2$. Keeping terms up to order $(\dfrac{\hat{q}\cdot \bm{\alpha}}{\omega})^4$ and solving the quadratic equation, the dispersion relation is obtained as 
\begin{equation}
\omega^{AP}(q)= |\hat{q} \cdot \bm{\alpha}| \sqrt{a_1+(1+a_2) |q|^2}.
\end{equation}
\section{Dynamical chemical potential imbalance} \label{appb}
The dynamical charge imbalance between the bulk and the boundaries is given by
\begin{equation}
\dfrac{d \tilde{n}{(\bm{r},t)}}{d t}=\dfrac{e^2}{2 \pi^2} \bm{\tilde{E}}{(\bm{r},t)} \cdot \bm{{\cal B}}^{el},
\end{equation}
which leads to 
\begin{equation}
\tilde{n}{(\bm{r},t)}=\dfrac{e^2}{2 \pi^2}\tau_a \bm{\tilde{E}}{(\bm{r},t)} \cdot \bm{{\cal B}}^{el}.
\end{equation}
Using the 3D electron density, i.e. $n=\dfrac{\mu^3}{3 \hbar v_F^3}=n_0+\tilde{n}{(\bm{r},t)}$, the shift in the chemical potential would be
\begin{equation}
\mu^3=\mu_0^3+\tilde{\mu}_{(\bm{r},t)}^3 \ \ \ \longrightarrow \ \ \mu= \mu_0 (1+(\dfrac{\tilde{\mu}_{(\bm{r},t)}}{\mu_0})^3)^{1/3},
\end{equation}
where
\begin{equation}
\tilde{\mu}{(\bm{r},t)}=(\dfrac{3 e^2}{2} \tau_a \bm{\tilde{E}}{(\bm{r},t)} \cdot \bm{{\cal B}}^{el})^{1/3}.
\end{equation}
Assuming $\mu{(r,t)} \ll \mu_0$, the shift in the chemical potential is simplified 
\begin{equation}
\delta \mu^{(2)}{(r,t)}=\mu - \mu_0 \simeq \dfrac{e^2}{2 \mu^2} \tau_a \bm{\tilde{E}}{(\bm{r},t)} \cdot \bm{{\cal B}}^{el}=\dfrac{e^2}{2 \mu^2} \tau_a \bm{\tilde{E}} \cdot \bm{{\cal B}}^{el} e^{i(\bm{q}\cdot \bm{r}-\omega t)}.
\end{equation}
The distribution function is defined as $\delta f_\chi (\bm{k},\bm{r},t)=(-\dfrac{\partial f^{(eq)}}{\partial \epsilon_k})\delta \mu^{(2)}{(\bm{r},t)} $, which satisfies $\delta {\tilde n}{(r,t)}=\sum_k {\cal D}_{(\hat{k})} \delta f_\chi (\bm{k},\bm{r},t)=\dfrac{e^2}{2 \pi^2}\tau_a \bm{\tilde{E}}{(\bm{r},t)} \cdot \bm{{\cal B}}^{el}$ up to the linear order of $\bm{{\cal B}}^{el}$.
\section{Thermal properties} \label{appc}
The total energy carried by the collective mode would be
\begin{equation}
{\cal U}=\sum_{\bm{q}} \omega_{q} \ {\cal G}^{(0)}(\bm{q},\bm{r},t),
\end{equation}
where ${\cal G}^{(0)}(\bm{q},\bm{r},t)=(e^{\beta \omega_{q}}-1)^{-1}$ is the equilibrium \textit{Bose-Einstein} distribution function and $\omega_q=|\hat{q} \cdot \bm{\alpha}| \sqrt{a_1+(1+a_2)|q|^2}$ represents the dispersion of the collective mode. The specific heat, i.e. ${\cal C}_v=\partial {\cal U}/\partial t$, can be obtained as
\begin{equation} \label{sp.heat}
{\cal C}_v(T)=\dfrac{\partial}{\partial T} \sum_{|\bm{q}|<\Lambda} \dfrac{\omega_q}{e^{\beta \omega_q}-1}=k_B \sum_{|\bm{q}|<\Lambda} \dfrac{\beta^2 \omega_q^2}{4 \sinh^2(\beta \omega_q/2)},
\end{equation}
where $\Lambda$ is an ultraviolet cutoff for the wave vector integrals. We assume that the pseudomagnetic field is parallel to the $z$-axis. The plasmon dispersion in the spherical coordinate is obtained as $\omega_q=\alpha \cos \theta \sqrt{a_1+(1+a_2)|q|^2}$, where $\theta$ is the angle between $\hat{z}$ and $\bm{q}$. With change of variables $x=\sqrt{a_1+(1+a_2)|q|^2}$ and $u=\cos \theta$, the renormalized specific heat in the spherical coordinate would be
\begin{equation} \label{sp-int}
\tilde{{\cal C}}_v=\dfrac{4 \pi^2 (1+a_2)^{3/2}}{k_B \Lambda^3}{\cal C}_v=(\dfrac{\theta_{AP}}{T})^2 \int_{-1}^{1} u^2 du \int_{\sqrt{a_1}}^{\xi} dx \dfrac{x^3 \sqrt{x^2-a_1}}{(e^{(\theta_{AP}/T)x u}-1)^2} e^{(\theta_{AP}/T)x u},
\end{equation}
where $\xi=\sqrt{a_1+a_2+1}$ and $\theta_{AP}=\alpha \Lambda/k_B$ is the corresponding Debye temperature for APs with $\alpha={\cal B}^{el}/2 k_F^2$. Therefore the specific heat behavior in terms of temperature can be obtained by numerical solution of the above integral.

To investigate the low temperature behavior ($T \ll \theta_{AP}$) of specific heat, let us initially consider this integral
\begin{equation}
A=\int_{-1}^{1} u^2 du \dfrac{e^{(\theta_{AP}/T)x u}}{(e^{(\theta_{AP}/T)x u}-1)^2} =\dfrac{1}{(b x)^3} \int_{-b x}^{b x} dt \dfrac{t^2 e^t}{(e^t-1)^2}
\end{equation}
with $b=\theta_{AP}/T$ and $t=bxu$. In the low temperature limit the variable $b$ goes to infinity ($b \rightarrow +\infty$), then the above integral becomes
\begin{equation}
A_{T\rightarrow 0} \simeq \dfrac{1}{(b x)^3} \int_{-\infty}^{+\infty} dt \dfrac{t^2 e^t}{(e^t-1)^2}=\dfrac{1}{(b x)^3} \dfrac{2\pi}{3}
\end{equation}
Substituting into Eq. (\ref{sp-int}) leads to
\begin{equation}
\tilde{{\cal C}}_v=\dfrac{2 \pi}{3} B(a_1,a_2) \dfrac{T}{\theta_{AP}} \ \ \ \ \ \ \ \ \ \ for \ \ \ \ \ T\ll \theta_{AP},
\end{equation}
where $B(a_1,a_2)=\int_{\sqrt{a_1}}^{\sqrt{a_1+a_2+1}} \sqrt{x^2-a_1} dx=\sqrt{(a_2+1)(a_1+a_2+1)}/2-\frac{a_1}{2}\ln|(\sqrt{a_2+1}+\sqrt{a_1+a_2+1})/\sqrt{a_1}|$.

In the high temperature limit, on the other hand, we use $e^{(\theta_{AP}/T) xu}=1+(\theta_{AP}/T) xu+...$ in Eq. (\ref{sp-int}) which gives
\begin{equation}
\tilde{{\cal C}}_v=\dfrac{2}{3} (a_2+1)^{3/2} \ \ \ \ \ \ \ \ \ \ for \ \ \ \ \ T \gg \theta_{AP}
\end{equation}

The thermal conductivity, $\kappa^{th}$, is defined as a coefficient of the heat current $\bm{j}^{th}=\kappa^{th} (- \nabla T)$.
The thermal current associated to the unidirectional AP mode in terms of its spectrum is provided by
\begin{equation} 
\bm{j}^{th}=\sum_{|\bm{q}| < \Lambda} \omega_q (\nabla_q \omega_q) \ \delta {\cal G}(\bm{q},\bm{r},t).
\end{equation}
where $\delta {\cal G}(\bm{q},\bm{r},t)$ is the deviation of distribution function due to the temperature gradient. The Boltzmann equation of bosonic distribution function with momentum $\bm{q}$ would be
\begin{equation}
\partial_t {\cal G}(\bm{q},\bm{r},t)+\dot{\bm{q}}\cdot \nabla_q {\cal G}(\bm{q},\bm{r},t)+\dot{\bm{r}}\cdot \nabla_r {\cal G}(\bm{q},\bm{r},t)=-\dfrac{ {\cal G}(\bm{q},\bm{r},t)- {\cal G}_0(\bm{q},\bm{r},t)}{\tau_p(T)}
\end{equation}
Using the stationary condition, i.e. $\partial_t {\cal G}(\bm{q},\bm{r},t)=0$, and $\dot{\bm{q}}=0$ we have
\begin{equation} \label{eq.boltz}
\delta {\cal G}(\bm{q},\bm{r},t)={\cal G}(\bm{q},\bm{r},t)- {\cal G}_0(\bm{q},\bm{r},t)=-\tau_p(T) \dot{\bm{r}}\cdot \nabla_r {\cal G}(\bm{q},\bm{r},t).
\end{equation}
It is straightforward to derive the following expressions;
\begin{equation}
\dot{\bm{r}}=\nabla_q \omega_q=\dfrac{\partial}{\partial q}[\alpha \sqrt{a_1+(1+a_2)|q|^2}]=\dfrac{\alpha^2 (1+a_2)}{\omega_q}|q|,
\end{equation}
and
\begin{equation}
\nabla_r {\cal G}(\bm{q},\bm{r},t)=\nabla T \dfrac{\partial {\cal G}(\bm{q},\bm{r},t)}{\partial T}.
\end{equation}
Substituting into Eq. (\ref{eq.boltz}), we may write
\begin{equation}
\delta {\cal G}(\bm{q},\bm{r},t)=- \tau_p(T) \dfrac{\alpha^2 (1+a_2)k_B \beta^2}{4 \sinh^2(\beta \omega_q/2)} |q| \nabla T
\end{equation}

Hence, the thermal current is given by
\begin{equation}
j^{th}=\tau_p(T)k_B (1+a_2)^2\alpha^4 \beta^2 \sum_{|\bm{q}| < \Lambda} \dfrac{q_z^2}{ 4\sinh^{2}(\beta \omega_q/2)} (-\nabla T)
\end{equation}
The renormalized thermal conductivity in the spherical coordinate is given by the following integral;
\begin{equation}
\begin{split}
\tilde{\kappa}^{th}=\dfrac{4 \pi^2}{\tau_p k_B (1+a_2)^{1/2} \Lambda^3}\kappa^{th}&=\alpha^2 b^2 \int_{\sqrt{a_1}}^{\xi} x (x^2-a_1)^{3/2} dx \int_{-1}^{1} \dfrac{e^{xub}}{(e^{xub}-1)^2} du \\ &
= \alpha^2 b \int_{\sqrt{a_1}}^{\xi} dx [\dfrac{(x^2-a_1)^{3/2}}{e^{-xb}-1}-\dfrac{(x^2-a_1)^{3/2}}{e^{xb}-1}]
\end{split}
\end{equation}
In the low temperature limit or $b \rightarrow +\infty$ we have
\begin{equation}
\tilde{\kappa}^{th}=\alpha^2 b \ C(a_1,a_2)|_{\sqrt{a_1}}^{\sqrt{a_1+a_2+1}}
\end{equation}
where $C(a_1,a_2)= \dfrac{1}{8} \sqrt{x^2-a_1}(5 a_1 x-2 x^3)-\dfrac{3}{8} a_1^2 \ln(x+\sqrt{x^2-a_1})$. Therefore, the low temperature behavior of $\tilde{\kappa}^{th}$ is proportional to $\tilde{\kappa}^{th} \propto ({\cal B}^{el})^2 (\theta_{AP}/T)$. It is straightforward to prove that the thermal conductivity in the high temperature limit becomes temperature independent and proportional to $({\cal B}^{el})^2$, i.e. $\kappa^{th} \propto ({\cal B}^{el})^2 $.

\section{Anomaly induced collective excitations in Weyl semimetals} \label{appd}

\begin{table}[ht]  
\caption{The physical interpretation of various collective modes in a Weyl semimetal in the presence and absence of external fields.} \label{table1}  
\centering 
\begin{tabular}{| c | c | c | c |} 
\hline \hline  
{\bf Collective mode} & {\bf \specialcell{ Coulomb \\[1 ex] interaction}} &  {\bf \specialcell{External \\[1 ex] fields}} & {\bf Physical interpretation}  \\ [0.5ex] 
\hline
Anomalous zero sound \citep{Gorsky_2013} & off & $\bm{B}$ & \specialcell{Linear dispersion: $\omega=c_s^{\pm} q$, $c_s^{\pm} \propto \pm B, \sqrt{F_0}$. \\[1 ex] Magnetic field lifts the degeneracy  of normal zero sound \\[0.1 ex] into two + and - branches. } \\ [1 ex]
\hline
Zero sound \citep{PhysRevB.98.165122} & off & \specialcell{$\bm{E}$ \\[1 ex] $\bm{B}$} & \specialcell{Topological effect changes the instability conditions of \\[0.1 ex] normal zero sound. \\[1 ex] It leads to the Landau damping even in the region where \\[0.1 ex] normal zero sound is undamped.}\\ [1 ex]
\hline
Chiral magnetic plasmon \citep{Gorbar_2017}& off & \specialcell{$\bm{B^{el}}$ \\[1 ex] $\bm{B}$} & \specialcell{Its origin is the fluctuations of both charge and chiral \\ [0.1 ex] current densities. \\ [1 ex] The plasma frequency is decomposed into \\ [0.1 ex] two branches under a magnetic field.} \\ [1 ex]
\hline
Chiral magnetic wave \citep{PhysRevD.91.125014}& off & $\bm{B}$ & \specialcell{Linear dispersion: $\omega=v_B q$, $v_B \propto \chi B$. \\[1 ex]
Its velocity does not depends on the detail of $\epsilon(p)$ or the \\ [0.1 ex] collision process.}\\ [1 ex]
\hline
Chiral sound wave \citep{PhysRevResearch.1.032040} &  off &  \specialcell{$\bm{E^{el}}$\\[1 ex] $\bm{B^{el}}$} & \specialcell{Linear dispersion: $\omega=v_s q$, $v_s \propto \hat{q} \cdot B^{el}$. \\[1 ex] It propagates along the $\bm{B^{el}}$ and origins from the chiral  \\[0.1 ex] anomaly with pseudofields.\\[1 ex]  It modifies the standard acoustic phonon dispersion.} 
\\[1 ex] 
\hline
Plasmon mode \citep{PhysRevB.91.035114} & on & \specialcell{$\bm{E}$ \\[1 ex] $\bm{B}$} & \ \ \specialcell{Its origin is the fluctuations of both the charge and chiral \\ [0.1 ex] current densities. \\ [1 ex] The chiral anomaly induces a Lifshitz transition to the plasmon \\ [0.1 ex]  frequency.} \\ [1 ex]
\hline
Chiral zero sound \citep{Song_2019} & on & $\bm{B}$ & \specialcell{Linear dispersion: $\omega=c_s q$, $c_s \propto \hat{q} \cdot \bm{B}$. \\ [1 ex] It emerges in a Weyl semimetal with at least two pairs of \\ [0.1 ex] Weyl points and propagates along the magnetic field. \\ [1 ex] 
It is also manifested to the unidirectional and unusual \\ [0.1 ex] thermal conductivity \citep{Song_2019,PhysRevX.9.031036}.} \\ [1 ex]
\hline
Chiral plasmon mode \citep{Song_2019} & on & $\bm{B}$ &  \specialcell{ Its origin is the inter-node fluctuations in the chiral limit. \\ [1 ex] 
The plasmon gap is proportional to the magnetic field, \\ [0.1 ex]  $\omega{(q \rightarrow 0)} \propto \hat{q} \cdot \bm{B}$  and propagates along the $\bm{B}$.  
}\\ [1 ex]
\hline
Anomalous plasmon mode & on &  \specialcell{$\bm{E^{el}}$\\[1 ex] $\bm{B^{el}}$} & \specialcell{It propagates along the $\bm{B^{el}}$ and origins from the local charge \\[0.1 ex] fluctuations between the bulk and the boundaries. \\[1 ex] The plasmon gap is proportional to the pseudomagnetic \\[0.1 ex] field, $\omega {(q\rightarrow 0)}\propto \hat{q} \cdot \bm{B}^{el}$.\\[1 ex] It leads to the unprecedented thermal conductivity along \\[0.1 ex] the $\bm{B}^{el}$ which violates the Widemann-Franz law.} \\ [1 ex]
\hline 
\end{tabular} 
\end{table}
\clearpage
\end{widetext}

\bibliography{ref}

\end{document}